\documentclass[aps,prb,twocolumn,showpacs,amsmath,amssymb,superscriptaddress]{revtex4-1}
\usepackage{graphicx}
\usepackage{bm}
\usepackage{graphicx}
\usepackage{epsfig}
\include{graphic}

\begin{document}

\title{On-chip artificial magnon-polariton device for voltage control of electromagnetically induced transparency}
\author{S. Kaur}
\affiliation{Department of Physics and Astronomy, University
of Manitoba, Winnipeg, Canada R3T 2N2}
\author{B. M. Yao}
\affiliation{Department of Physics and Astronomy, University
of Manitoba, Winnipeg, Canada R3T 2N2}
\affiliation{National Laboratory for Infrared Physics, Chinese Academy of Sciences, Shanghai 200083, People’s Republic of China}
\author{Y. S. Gui} \email{ysgui@physics.umanitoba.ca}
\author{C.-M. Hu} \email{hu@physics.umanitoba.ca; \\URL: http://www.physics.umanitoba.ca/$\sim$hu}
\affiliation{Department of Physics and Astronomy, University
of Manitoba, Winnipeg, Canada R3T 2N2}

\begin{abstract}

We demonstrate an on-chip device utilizing the concept of artificial cavity magnon-polariton (CMP) coupling between the microwave cavity mode and the dynamics of the artificial magnetism in a split ring resonator. This on-chip device allows the easy tuning of the artificial CMP gap by using a DC voltage signal, which enables tuneable electrodynamically induced transparency. The high tunability of the artificial magnon-polariton system not only enables the study of the characteristic phenomena associated with distinct coupling regimes, but also may open up avenues for designing novel microwave devices and ultra-sensitive sensors.
\end{abstract}

\maketitle
 
When an electromagnetic wave propagates in a magnetic material, its magnetic fields can drive the magnetization precession and the mutual coupling between the macroscopic electrodynamic and magnetization dynamic results in a hybrid electromagnetic mode of media, i.e., magnon polariton\cite{Mills1974}. In light of this principle, a cavity magnon polariton (CMP) has been recently studied in a coupled magnon-cavity photon system \cite{Bai2015}, where the general feature of the CMP is described in a concise classical model \cite{Bai2015} which accurately highlights the key physics of phase correlation between the cavity and magetization resonances. This general CMP model can be used to quantitatively analyze the characteristic features of magnon-photon coupling experiments recently performed by many different groups \cite{ Huebl2013, Tabuchi2014, Zhang2014, Goryachev2014, Bhoi2014, Bai2015, Haigh2015, Lambert2015, Abdurakhimov2015}, in which a low damping bulk ferromagnetic insulator is set either on-top of a superconducting co-plannar waveguide or inside a high quality 3D microwave cavity. 

The intriguing physics of CMP opens up new avenues for materials characterization and microwave applications. For example, CMP effect has recently been observed by setting miligrams of magnetic nano-particles inside a special circular waveguide cavity \cite{Yao2015}, where the CMP coupling can be analyzed in the simple 1D configuration by using either the straightforward transfer matrix method \cite{Yao2015}, or equivalently, the 1D scattering theory \cite{Cao2015}. It is found that the CMP coupling enables quantifying the complex permeability of magnetic nanoparticles with high sensitivity, which was an outstanding challenge for the biomedical applications of magnetic nanoparticles \cite{Yao2015}. In 3D microwave cavities, it is shown that the CMP coupling leads to the electromagnetically induced transparency (EIT), which is tuneable by applying an external magnetic field \cite{Zhang2014}. Such a tuneable EIT is of great importance for designing microwave circuits.

From the perspective of device application the external magnetic field used to alter the resonance frequency of the magnon, and the size of the 3D cavity are not favourable. Inspired by the study of artificial magnetism\cite{Pendry1999, Smith2004}, where the non-magnetic conducting material is designed to provide a magnetic response at microwave frequencies, in this paper, we report on a 2D artificial CMP system created by integrating an artificial magnetic resonator with an on-chip cavity resonator. Here the cavity mode and the artificial magnon mode are generated by a cut wire resonator connected to a transmission line and a varactor loaded split ring resonator mutually coupled with the cut wire respectively. The microwave magnetic field radiated by the cut wire resonator induces a microwave current in the split ring resonator and thus produces an effective magnetic response, which can be characterized by the resonant permeability of the structure.\cite{Pendry1999} By altering the DC voltage bias of the varactor diode, the resonance frequency of the split ring resonator can be varied in a wide range. Aligning the resonance frequency of the photon and the pseudo-magnon subsystem, the artificial CMP gap induced by electrodynamic coupling is observed, creating tunable EIT using a DC voltage signal. To explain the experimental observations, we have developed the coupled LCR circuit model, where the coupling strength is determined by the mutual inductance and is proportional to the distance between two structures based on Biot Savart's law. The device based on this 2D artificial CMP may create new scenarios to study the coupled photon-quasi-particle system, allowing the observation of characteristic phenomena on-chip associated with different coupling regimes including weak coupling, strong coupling, EIT and Purcell effect controlled by a DC voltage. 

Similar to that of the CMP\cite{Bai2015}, the dispersion of the artificial CMP as detailed in the discussion below can be completely described by five parameters: the resonance frequency of the cavity mode ($\omega_c$), the resonance frequency of the pseudo-magnon mode ($\omega_s$), the damping factor of the cavity mode ($\Delta\omega_c$), the damping factor of the pseudo-magnon mode  ($\Delta\omega_s$) and the coupling strength ($\kappa$) between the two modes. To quantify them from the observed microwave response of the artificial CMP the transmission line theory of the coupled LCR circuit is developed which bridges the gap between simple circuit analysis and Maxwell’s equations by describing wave propagation in terms of current, voltage and impedance\cite{Pozarbook}. 

\begin{figure} [t]
\begin{center}\
\epsfig{file=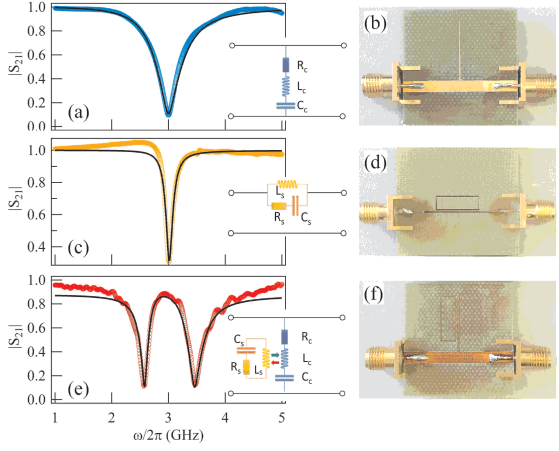,width= 7.5 cm} 
\caption{(color online). Experimentally measured (symbols) and analytically calculated (solid lines) transmission curves for (a) the cut wire resonator connected to a 50 $\Omega$ transmission line, (c) the split ring resonator mutually excited by 150 $\Omega$ transmission line, and (e) the coupled cut wire and split ring resonator separated by a distance of 2.7 mm. The linewidth of both resonator structures is 0.2 mm. The insets show the corresponding equivalent LCR circuits and (b), (d) and (f) are the photograph of the fabricated microwave devices. The S-parameter experiments were performed by connecting the device to a vector network analyzer. 
}\label{fig1}
\end{center}
\end{figure}

We first used the transmission model described in Ref. \onlinecite{Fu2007}, in which the cut wire resonator and the split ring resonator is directly excited by the incident electromagnetic radiation, to characterize the LCR parameter in the individual cut wire resonator and  split ring resonator, respectively. As shown in the inset of Fig. \ref{fig1}(a) and (c) the cut wire resonator and the split ring resonator can be treated as the LCR circuits and the deduced impedances are  $Z_{c} = R_{c} + i(\omega L_{c} - 1/\omega C_{c})$ and $Z_{s} = (i\omega L_{s} - R_{s}\omega^{2}L_{s}C_{s})/(1-\omega^{2}L_{s}C_{s}+i\omega C_{s}R_{s})$ where the subscripts "c" and "s" denote the parameter for the cut wire resonator and the split ring resonator respectively, and $\omega/2\pi$ is the microwave frequency. Through the transfer matrix \cite{Pozarbook} the related transmission coefficient ($S_{21}$ parameter) can be easily obtained as $S_{21}=2Z_c/(2Z_c+Z_0)$ and $S_{21}=2Z_0/(2Z_0+Z_s)$ for the cut wire resonator and the split ring resonator respectively, where  $Z_{0}$ is the characteristic impedance of the transmission line. 
 
When combining the two resonators to observe the artificial CMP, the LCR circuit shown in the inset of Fig. \ref{fig1}(e) was used, where the interaction between two resonators is described by mutual inductance $M$. The $S_{21}$ parameter for this configuration is obtained by solving for the net impedance of the circuit and using the relationship between the transmission parameter and the transfer matrix. After simplification, the $S_{21}$ parameter can be written as

\begin{equation}
S_{21} = 1-\dfrac{\dfrac{Z_0}{Z_0+2R_c}\Delta\omega_c }{i(\omega-\omega_{c})+\Delta\omega_{c}+\dfrac{\kappa^{2}}{i(\omega-\omega_{s})+\Delta\omega_{s}}}
\label{EIT}
\end{equation}

\noindent where $\omega_{c}=1/\sqrt{L_{c}C_{c}}$ and $\omega_{s}=1/\sqrt{L_{s}C_{s}}$ are the resonance frequencies of the cavity mode and the pseudo-magnon mode respectively, $\Delta\omega_{c}=\dfrac{C_{c}\omega \omega_{c}^{2}(R_{c}+Z_{0}/2)}{\omega+\omega_{c}} $ and $\Delta\omega_{s}=\dfrac{C_{s}R_{s}\omega_{s}^{2}\omega}{\omega+\omega_{s}}$ are the damping factors related to the cavity mode and the pseudo-magnon mode respectively and $\kappa=\omega^2 M \omega_{c} \omega_{s}\sqrt{{C_{c}C_{s}}/(\omega+\omega_{c})(\omega+\omega_{s})}$ is a measure of the coupling strength between the two modes and is linearly dependent on the mutual inductance $M$. In the vicinity of the polariton gap which occurs at the condition of $\omega\sim\omega_c\sim\omega_s$, it can be found that $\Delta\omega_{c}\sim(R_c+Z_0/2)/2L_c$, $\Delta\omega_{s}\sim{}R_s/2L_s$, and $\kappa\sim{}M\omega/2\sqrt{L_cL_s}$.  Notice that Eq. (\ref{EIT}) has a similar expression as that for the CMP due to coupling between photon and magnon, indicating the similar physical origin of the artificial CMP\cite{Bai2015}. 

In order to validate this coupling model based on the transmission line theory, the cut wire resonator,  the split ring resonator, and their combination shown in Fig. \ref{fig1} (b), (d), and (f) were fabricated on 1.55 mm thick FR4 substrate with a loss tangent of 0.017 using the LPKF S103 circuit board plotters. The copper thickness is 35 $\mu$m. The cut wire resonator is positioned at the center of a 50 $\Omega$ microstrip line, the length and width of cut wire is 15.1 mm and 0.2 mm, respectively. The dimension of split ring resonator is 4 mm in width and 12 mm in length with a linewidth of 0.2 mm. Along the width of the split ring resonator, a 0.2 mm gap is located in the center and a 0.2 mm wide 150 $\Omega$ transmission line is used to excite the resonance.  In Fig. \ref{fig1}(f) the separation between the center of the two resonator structure is 2.7 mm and the microstrip line has a characteristic impedance of 50 $\Omega$. The transmission parameter of these fabricated devices was then measured using an Agilent vector network analyzer N5230C. 

The experimentally measured transmission curves are shown in Fig. \ref{fig1}(a), (c), and (e) respectively where a transmission minimum is observed at the resonance frequency of 3.0 GHz in the spectrum for both the resonators whereas a transmission maximum is obtained at 3 GHz when a combination of them is used. This appearance of a narrow transmission window centred at the resonance frequency of the individual resonator corresponds to the EIT phenomenon. EIT was originally observed in atomic gases\cite{Hau1999} due to destructive interference between two different excitation pathways\cite{Harris1990}, and soon was realized in its classical analogue in metamaterials \cite{Fedotov2007, Papasimakis2009, Tassin2009, Liu2009, Liu2011} as well as recently in the CMP system\cite{Zhang2014}.

The transmission curves obtained for the cut wire resonator and the split ring resonator using our transmission line model are also displayed in Fig. \ref{fig1}(a) and (c) respectively with the corresponding values of $R_c$=($2.87 \pm 0.07$) $\Omega$, $C_c$=($0.499 \pm 0.001$) pF and $L_c$ ($5.64 \pm 0.01$) nH  for the cut wire resonator, and $R_s$=($0.091 \pm 0.003$) $\Omega$ , $C_s$=($8.78 \pm 0.09$) pF  and $L_s$=($0.314 \pm 0.002$) nH for the split ring resonator. Hence the important parameters of $\omega_c$, $\omega_s$, $\Delta\omega_c$ and $\Delta\omega_s$ are determined. Then the measured transmission spectrum corresponding to the coupled resonator(shown in Fig. \ref{fig1}(e)) was calculated with just one fitting parameter i.e. mutual inductance ($M$ = 0.74 nH).

\begin{figure} [t]
\begin{center}
\epsfig{file=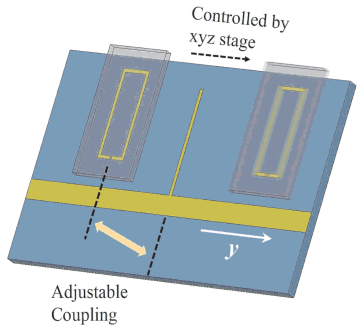,width= 5 cm} 
\caption{(color online). Schematic diagram of the experimental setup used to achieve active tunability of EIT. The 50$\Omega$ transmission line integrated with a cut wire resonator in the center is directly connected to the vector network analyzer connectors whereas the chip of the split ring resonator is mounted on the Velmex x-y-z stage which is used to tune the distance between two resonators.
}\label{fig2}
\end{center}
\end{figure}

Since the mutual inductance $M$ determines the coupling $\kappa$, which characterizes the energy transfer efficiency from the the cavity system to the quasi-particle system\cite{Bai2015}, it is necessary to understand its physical origin in an artificial CMP for microwave applications. In the second experiment, the distance between two resonators fabricated as different chips was changed in the horizontal direction using an x-y-z stage. Figure \ref{fig2} shows the schematic diagram of the experimental setup used where the cut wire resonator shown was directly connected to the vector network analyzer whereas the split ring resonator was mounted on the x-y-z stage and the separation between the two chips was less than 20 $\mu$m. 

\begin{figure} [t]
\begin{center}
\epsfig{file=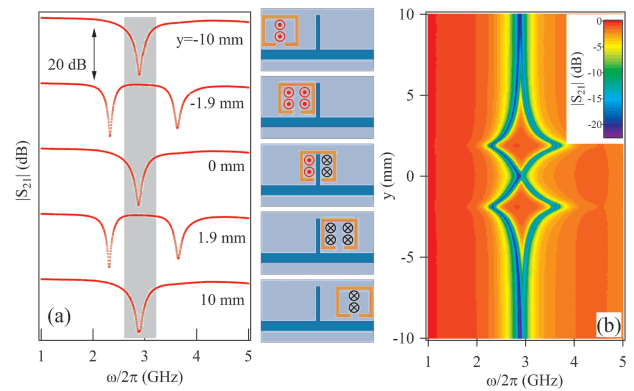,width= 8.5 cm} 
\caption{(color online). (a) The left panel shows some typical transmission spectra (offset for clarity) obtained at different distances between two resonators and the right panel shows the schematic diagram of the corresponding measurement configuration along with the amount of flux flowing through the  split ring resonator due to the microwave current in the cut wire resonator. (b) The transmission amplitude mapping result for distance $y$ between two resonators as a function of frequency.
}\label{fig3}
\end{center}
\end{figure}

The left panel of Fig. \ref{fig3}(a) shows some typical $S_{21}$ spectra obtained at different values of distance $y$ between the cut wire resonator and the center of the split ring resonator, and the right panel shows the corresponding amount of magnetic flux passing through the split ring resonator due to the current induced in the cut wire resonator by the electromagnetic field of the microwave flowing in the transmission line. From Fig. \ref{fig3}(a) it can be easily seen that as the distance $y$ decreases, the magnetic flux through the split ring resonator increases which corresponds to an increase in the coupling strength $\kappa$. Due to this increased coupling strength, the transmission window related to EIT increases and reaches a maximum when two resonators are closest without any overlap. However, when the cut wire resonator is exactly at the centre of the split ring resonator ($y=0$), the induced current in the cut wire resonator generates electromagnetic fields that induce equal and opposite currents in the two arms of the split ring resonator which exactly cancel each other. Hence, the EIT effect vanishes at the centre but reappears as soon as the centres of the two resonators are no longer aligned. 

\begin{figure} [t]
\begin{center}
\epsfig{file=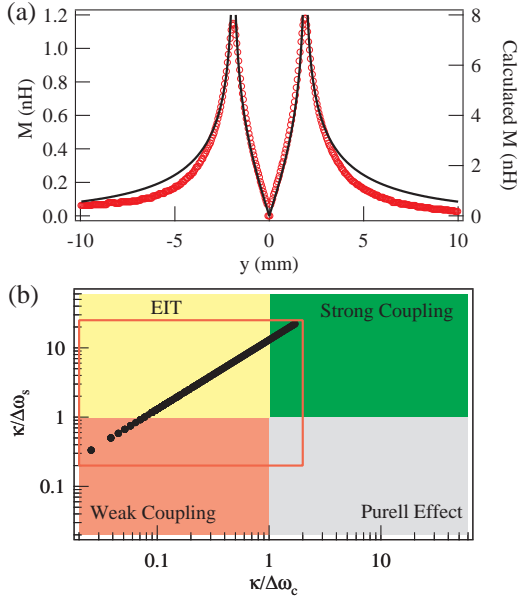,width= 7 cm} 
\caption{(color online). (a)Experimental (symbol) and theoretical (solid line) mutual inductance curve as a function of distance $y$ between two resonators. (b) Different coupling regimes separated by $\kappa/\Delta\omega_c$ and $\kappa/\Delta\omega_s$. The circles are experimental data for coupled system with different coupling strength determined by $y$.  
}\label{fig4}
\end{center}
\end{figure}

Figure \ref{fig3}(b) shows the transmission amplitude mapping result obtained by tuning the distance $y$ in the horizontal direction. Based on the amplitude mapping result, a relationship between the mutual inductance $M$ and the distance $y$ can be established and the result is shown in Fig. \ref{fig4}(a). The mutual inductance $M$ follows the same pattern with distance as the transparency window associated with EIT (shown in Fig. \ref{fig3}(b)) which validates our transmission line model in which the coupling strength is shown to be linearly dependent on the mutual inductance. This experimentally determined relationship between the mutual inductance and the distance can be theoretically explained by determining the magnetic flux ($\phi$) through the split ring resonator due to the induced current ($I$) flowing in the cut wire resonator, given by

 \begin{equation}
 M = \dfrac{\phi}{I} = \dfrac{\oint \mathbf{B}\cdot d\mathbf{A}}{I}
 \end{equation} 

\noindent where $\mathbf{B}$ is the microwave magnetic field produced by the cut wire resonator. Since the cut wire resonator is of finite length then the resulting magnetic field is $B = \dfrac{\mu_{0}}{4\pi x}(\cos\theta_{1}+\cos\theta_{2})$ where $\cos\theta_{1}$ and $\cos\theta_{2}$ are the angles subtended by the two end points of the cut wire resonator at a point located at a distance $x$ from it. The net flux and the mutual inductance between two resonators can then be estimated by integrating the magnetic field over the area of the split ring resonator. For simplicity by neglecting the contribution caused by the substrate here we integrate the magnetic field over the width of the split ring resonator and simply multiply with its length and the result is shown in Fig. \ref{fig4}(a) where good qualitative agreement is observed on the distance dependence between the experimental result and the theoretical expectation. 

We note that different coupling regimes can be characterized by the relative strength of $\kappa/\Delta\omega_c$ and $\kappa/\Delta\omega_s$ as shown in Fig. \ref{fig4}(b), which reflects the energy exchange between the photon and pseudo-magnon sub-systems and the energy dissipation in them. By tuning the coupling strength $\kappa$ through the distance, the presented experiments (symbols) cover the regime of weak coupling, EIT and strong coupling. We can further introduce additional tunable freedom using a loaded resistance, which can effectively increase the damping parameter of the resonator\cite{Tassin2009, Zhang2012, Sun2014}. Therefore, the tunable regime based on the presented structure is bounded by the rectangular shape (red line) shown in Fig. \ref{fig4}(b), which clearly demonstrates the high tunability of our compact artificial CMP device, and allows the coupling phenomena to be observed in distinct regimes. 

\begin{figure} [t]
\begin{center}
\epsfig{file=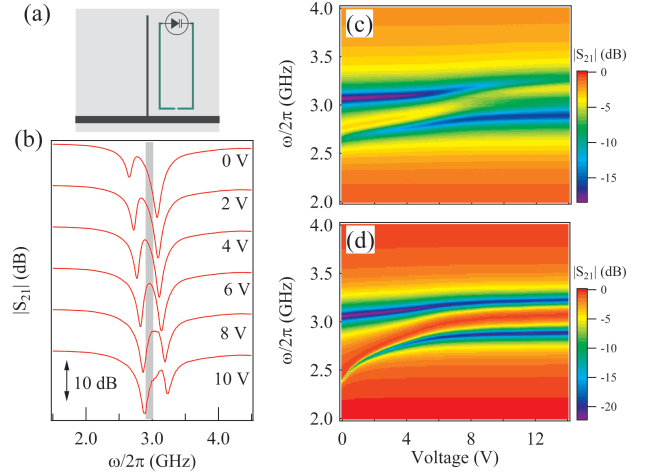,width= 8.5 cm} 
\caption{(color online). (a) Schematic diagram of the artificial CMP device with a distance of about 4 mm between two resonators. (b) Typical $S_{21}$ spectra (offset for clarity) obtained at different values of applied voltage. At about 3.0 GHz, the device changes from being transparent to opaque as the voltage is changed. (c) Measured and (d) calculated  transmission dispersion of artificial CMP formed using a varactor diode biased by DC voltage signals.}\label{fig5}
\end{center}
\end{figure}

In the third experiment, an on-chip artificial CMP device has been fabricated and demonstrates the capability of the voltage control of EIT.  Based on the dependence of mutual inductance on the distance between two resonators, the appropriate coupling region was chosen to demonstrate this active voltage control of the transparency window associated with EIT, where the artificial CMP gap is generated by electrodynamic coupling.  In the device a varactor diode (Skyworks SMV 2019) was soldered onto the split ring resonator which had a capacitance tuning range from 2.2 pF to 0.3 pF\cite{Skyworks} as shown in the schematic diagram Fig. \ref{fig5}(a). As the DC voltage bias is tuned, the capacitance of the varactor loaded split ring resonator changes. Consequently, the resonance frequency $\omega_s$ of the split ring resonator  will change while the resonance frequency $\omega_c$ of the cut wire resonator remains constant at 3 GHz.

In the device configuration shown in Fig. \ref{fig5}(a) the split ring resonator cannot be directly excited by the 50 $\Omega$ transmission line, and has been proven by additional experiments (not shown). Due to the mutual coupling between the photon and pseudo-magnon subsystem, both resonances have been observed even at conditions of $\omega_s\neq\omega_c$. As shown in the typical $S_{21}$ spectra in Fig. \ref{fig5}(b) the amplitude of the resonance of the split ring resonator is an order of magnitude weaker compared with that of the cut wire resonator when $\omega_s$ is far from $\omega_c$, indicating the less-effective efficiency of the energy transfer from the cut wire resonator to the split ring resonator. The amplitude of the resonance of the split ring resonator is significantly enhanced when $\omega_s$ approaches $\omega_c$ due to the fact that the coupling of pseudo-magnon and cavity mode hybridizes them and generates the artificial CMP. A general phenomenon known as "avoided level crossing" occurs when the two subsystem have identical resonance, i.e. $\omega_c/2\pi\approx\omega_s/2\pi\approx3.0$ GHz at the external voltage of about 6.5 V, where the photon-like and the pseudo-magnon-like polariton have equal amplitude and neither of them occurs at 3.0 GHz. As a result the polariton gap (about 0.3 GHz) of the artificial CMP is observed, creating EIT near 3 GHz.

Figure \ref{fig5}(c) maps the measured dispersion of an artificial CMP formed by using voltage-controlled microwave device, where the hybridization of the pseudo-magnon and cavity modes is clearly seen, generating the photo-like and pseudo-magnon-like polaritons separated by a gap near 3.0 GHz. As the applied voltage increases, the device changes from being partially transparent to completely transparent and then to being completely opaque at about 3.0 GHz. One can calculate the dispersion of an artificial CMP based on Eq. (\ref{EIT}) by integrating the equivalent circuit of the varactor diode, which includes multiple electronic elements. Here we highlight the electrically tunable feature of this hybrid device by assuming that only the capacitance varies with the external applied DC voltage. Assuming $C_v$ of the varactor is in series connection with $C_s$= 2.7 pF for the varactor loaded split ring resonator with $L_s$ = 0.88 nH, the $S_{21}$ spectra is calculated as shown in Fig. \ref{fig5}(d), qualitatively reproducing the observed EIT feature in the coupled photon-pseudo-magnon system. While tunable EIT has been achieved by the CMP controlled by a static magnetic field, the artificial CMP is an on-chip device controlled by the DC voltage and therefore is more convenient for practical applications.

In summary, by introducing the new concept of artificial cavity magnon polariton, an on-chip hybrid device has been fabricated, where the artificial CMP gap induced by electrodynamic coupling is demonstrated, achieving voltage-tuneable electrodynamically induced transparency in microwave range. A transmission line model which describes the electrodynamic coupling between the cavity and the pseudo-magnon modes has been developed. Our 2D system has demonstrated high tunablility, allowing the observation and evaluation of characteristic phenomena associated with distinct coupling regimes. This work not only explores the physical understanding of the coupled dynamic system, but also could be potentially used in fabricating dynamic filters and switch devices.

This work has been funded by NSERC, CFI, the National Key Basic Research Program of China (2011CB925604), and the National Natural Science Foundation of China Grant No. 11429401. We would like to thank L. H. Bai, B. W. Southern, J. Dietrich, G. E. Bridges and L. Fu for discussions.

\end{document}